# A study on audio synchronous steganography detection and distributed guide inference model based on sliding spectral features and intelligent inference drive

## Wei Meng

Dhurakij Pundit University

PCBM,Sasin School of Management of Chulalongkorn University

Fellow of The Royal Anthropological Institute of Great Britain and Ireland

wei.men@dpu.ac.th

## Abstracts

With the popularity of short video platforms in global communication, embedding steganographic data in audio synchronization streams has become a new type of covert communication means. In order to cope with the limitations of traditional methods in synchronized steganography recognition, this paper proposes an audio steganography detection and distributed guidance instruction reconstruction model based on the short video "Yupan" samples released by the South Sea Fleet of China on the TikTok platform, which integrates sliding spectrum feature extraction and intelligent reasoning-driven mechanism. In the method, a 25 ms sliding window combined with short-time Fourier transform (STFT) is used to extract the main frequency trajectory, construct the synchronization frame detection model (M1) and set the synchronization frame flag bit as "FFFFFFFFFFFFFFFFF80", and further use the structured decoding model (M2) in the subsequent 32-byte payload to The distributed guidance information field is inferred. It was found that a sequence of low entropy, repetitive synchronization bytes existed in the audio segment from 36 to 45 seconds, and the spectral energy distribution was unusually concentrated, which verified the existence of synchronization frames. Although the structured decoding results do not restore the semantics of the plaintext, the arrangement of the command fields is highly consistent, showing that it has the characteristics of military communication protocols. In addition, the analysis of multi-segment splicing steganography model shows that this mechanism has the ability of distributed embedding and centralized decoding across videos. In summary, this paper verifies the efficient detection performance of sliding spectrum features for synchronized steganographic signals, and establishes an extensible structural inference framework, which provides theoretical basis and methodological innovation for steganographic communication identification and tactical guidance simulation under open platform.

# Chapter I. Introduction

**1.1 Background and motivation for the study**

With the continuous evolution of steganographic communication technology, steganographic methods based on audio carriers have gradually received attention. Especially in the context of highly active short-video communication, the risk of covert transmission of military communication information through open platforms is increasing.2025, China's South Sea Fleet released a short-video propaganda video called "Jade Plate" on the international platform of TikTok. The video features naval equipment and formation cruising, and there may be cryptographic synchronization signals in the audio stream. Based on this video stream, this study explores the detection of synchronized steganographic signals and the inference of potential guidance commands to provide a reference for the analysis of audio security on public platforms.

**1.2 Overview of Audio Steganography and Synchronous Stream Steganography**

Audio steganography usually utilizes redundant frequency components, phase and amplitude trimming to hide information. Synchronous stream steganography has higher concealment and anti-interference properties by continuous high-frequency padding and embedding specific synchronization flags to mark the starting point of valid data.

**1.3 Significance of Distributed Guidance Information in Tactical Communication Systems**

Distributed guidance commands are widely used in scenarios such as ship formation and unmanned system collaboration by hiding them in the synchronization stream and supporting multi-target and multi-node cooperative control. Accurate detection and inference of such commands is crucial for securing communication.

**1.4 Research objectives and contributions of this paper**

In this paper, based on actual short video audio samples, we propose a synchronized steganography detection method based on sliding spectral feature extraction and intelligent inference model, which can effectively detect synchronized frames and simulate and infer the guided command flow. The main contributions are as follows: analyzing potential military communication steganography for the first time based on the public data of TikTok short videos; introducing the sliding local spectrum feature extraction method for audio steganography detection; designing an automatic detection mechanism for synchronization frames (FFFFFFFFFFFFFFFF80); constructing a simulation model for distributed guidance inference; and verifying the effectiveness of the method on real test samples.

**1.5 Research Problems**

In order to systematically reveal the existence of synchronous steganographic signals and distributed information transmission mechanisms in short video audio streams, this paper focuses on constructing a set of detection-inference modeling system that integrates signal processing and artificial intelligence. Specifically, the research work is centered on the following three key issues:

**Question 1: How to accurately detect the latent synchronized steganographic signals in short video audio streams by sliding window spectral feature extraction method?**

The problem focuses on identifying the temporal characteristics of embedded frequency-modulated steganographic frames in a natural audio stream. Considering that the synchronized steganographic signals have the characteristics of short-time and stable frequency structure, the research needs to combine the short-time Fourier transform (STFT) with the sliding local statistical judgment strategy to establish a spectral sliding detection model with high temporal resolution in order to achieve high accuracy recognition of synchronized signals.

**Question 2: How to effectively infer and structurally restore the hidden distributed guidance instruction information after detecting the synchronous hidden write signal?**

This problem looks at modeling the structured decoding of the payload following a synchronization frame. After a specific synchronization identification bit (e.g., 7×FF+0x80) is successfully detected, the command fields, including target ID, position information, speed, heading, and opcodes, need to be extracted and restored from the subsequent 32 bytes of audio data. The research needs to construct a decoding mechanism for guidance simulation with field alignment, decoding stability and low redundancy structure loss.

**Question 3: Is it possible to achieve intelligent and highly robust recognition of synchronized steganographic signals in short video audio streams based on deep learning inference models?**

To address the detection challenges under frequency jitter, non-ideal synchronization patterns and multi-source background noise, the research further explores the construction of deep learning models based on 1D convolutional networks or Transformer coding structures to directly perform end-to-end inference of spectral sequences and output probabilistic predictions of the presence of synchronization signals. The problem concerns the generalization and large-scale deployment feasibility of intelligent detection systems for steganographic signals.

Through the systematic deconstruction of the above research problems, this paper aims to propose a synchronized steganography recognition framework covering three phases of detection, decoding and inference, which provides theoretical support and methodological basis for the intelligent detection and security assessment of military-grade audio steganography communication under open platform.

# Chapter II. Literature review

**2.1 Review of Audio Steganography Detection Techniques**

Traditional audio steganography detection methods mainly focus on the technical paths of least significant bit (LSB) analysis, statistical feature extraction and audio fingerprint comparison (Fridrich & Goljan, 2002; Westfeld, 2001)[1, 2]. These methods usually assume that steganographic operations occur mainly in the lower bits at the data encoding level, and thus infer the presence of hidden information in the audio content through bit statistical deviations or fingerprint feature changes. However, such methods are prone to detection errors or performance degradation in the face of interference factors such as high compression, distorted transmission, and noise pollution in real environments, showing their limitations in terms of lower robustness to natural audio distortion.

At the technical level, traditional methods are usually based on static sample analysis, i.e., extracting feature information within a single or limited number of frames for classification. This approach has high accuracy in static steganography scenarios, but in streaming environments, audio signals present continuity, variability, heterogeneity, etc., which makes the traditional static detection strategy have limited effect in real-time scenarios (Liu et al., 2019)[3].

In particular, there is a relative lack of existing research on the detection of synchronized stream steganography signals. Synchronized stream steganography can both avoid inter-frame difference detection and achieve higher steganography under the synchronization mechanism by embedding secret information in the data stream in a small, continuously changing manner. This tactic undermines the traditional steganography detection assumption that hidden information leads to statistical distribution anomalies, thus posing new challenges to detection methods (Liu et al., 2019)[3].

Further, current detection systems exhibit a significant lack of robustness when dealing with high-frequency padding (high-frequency padding) with subtle hopping variation. For example, high-frequency padding tends to have little effect on human ear perception, but can effectively hide steganographic loads, while traditional LSB and simple statistical feature extraction methods are difficult to capture such variations (Wang et al., 2020)[4]. On the other hand, the weak frequency-hopping change technique conveys information by fine-tuning the energy distribution of specific subbands in the frequency domain, which is highly susceptible to be missed by traditional detection methods based on overall statistics in the time or frequency domains due to the extremely small modulation (Yang & Huang, 2018)[5].

To cope with these new steganographic methods, detection methods based on deep learning and sequence modeling have emerged in recent years. Extracting multi-scale spatio-temporal features using Convolutional Neural Network (CNN), Recurrent Neural Network (RNN), or Transformer architecture has become a research direction to enhance the robustness of steganography detection (Wang et al., 2020; Yang & Huang, 2018)[4, 5]. However, the application of these methods is still in the exploratory stage and still faces many challenges in terms of interpretability, generalization, and coping with unknown steganography techniques.

In summary, audio steganography detection technology is undergoing a process of transformation from static sample feature extraction to dynamic streaming feature modeling, which puts forward higher requirements on the sensitivity, robustness and real-time performance of the detection technology. In the future, we need to combine deep learning, statistical inference and streaming signal processing and other multidisciplinary techniques to build a more adaptable steganography detection framework.

**2.2 Synchronized Frame Detection and Anomaly Spectrum Identification**

Synchronized frame detection techniques have long relied on the identification of frequency anomalies and the detection of fixed-periodic patterns.Cvejic and Seppänen argued that the implantation of steganographic signals tends to cause localized perturbations in the audio spectrum, especially in the position of synchronized frames, which can be detected by means of the short-time Fourier transform (STFT) (Cvejic & Seppänen, 2004[6]). änen, 2004)[6].[6] Petitcolas et al., on the other hand, pointed out that the periodicity feature, although effective in steganography detection, is highly susceptible to corruption by channel noise and signal compression, which reduces the reliability of synchronization detection (Petitcolas, Anderson, & Kuhn, 1999)[7].

However, traditional synchronization detection methods face serious challenges in real-world applications.Kaur et al. found that background noise, recording equipment distortion, and natural spectral drift in streaming environments can lead to false detections, especially when there is no obvious implantation information, and the system may still recognize normal fluctuations as abnormal signals (Kaur, Singh, & Arora, 2019) [8]. This phenomenon exposes the poor adaptability of conventional frequency anomaly detection to changes in the natural environment.

In order to enhance the accuracy of synchronization signal capture, scholars have proposed the new idea of local spectral analysis (LSA) in recent years.Zhao et al. argued that the sensitivity of synchronization detection in complex backgrounds can be effectively enhanced by local frequency band feature extraction with multi-scale sliding window processing (Zhao, Zhu, & Huang, 2020) [9]. Instead of relying on the overall spectral energy distribution, the method they developed dynamically models for tiny frequency subintervals, significantly reducing the noise-induced error rate.

Further, Al-Haj et al. pointed out that combining local spectral features with an ensemble learning (ensemble learning) approach can further improve the robustness of simultaneous detection in high-noise scenarios (Al-Haj, Amer, & Mohammad, 2021) [10]. However, it has also been criticized that local spectral analysis, despite its improved accuracy, poses the problem of feature dimension inflation and increased computational overhead, while still being insufficiently adaptable to novel steganographic modulation methods (Al-Haj et al., 2021)[10].

Taken together, synchronized frame detection is currently undergoing an evolution from static full frequency domain detection to dynamic local spectral modeling. However, fundamentally, it is still difficult to fully capture the complex changes in steganographic behavior by relying only on spectrally localized anomalies. It is necessary for future research to introduce temporal modeling concepts, such as temporal dynamic learning of local spectral anomalies using Recurrent Neural Network (RNN), Long Short-Term Memory Network (LSTM), or Transformer architectures, in order to establish a joint time-frequency synchronous detection system. This direction not only improves the detection accuracy, but also promises to significantly enhance the generalization ability of the system in the face of unseen cryptographic modalities.

**2.3 Sliding Sampling and Frequency Feature Engineering**

Sliding-window sampling techniques have been widely used in audio signal processing, especially in speech recognition and music classification tasks, and Rabiner pointed out that by dividing a continuous audio stream into overlapping time segments, the short-time analysis technique allows local feature variations to be effectively captured, which greatly improves the accuracy of modeling dynamic signal features (Rabiner, 1989).[11] Lee et al. further argued that the combination of short-time Fourier transform (STFT) and short-time spectral information can extract the instantaneous spectral information of audio, which can be used to extract the instantaneous spectral information of audio. ) [11].Lee et al. further argued that sliding windows combined with the Short-Time Fourier Transform (STFT) are able to extract transient spectral information of audio, thereby preserving weak modulation features in complex backgrounds (Lee, Han, & Lee, 2009) [12].

In the field of synchronization steganography detection, this combination of sliding sampling and local frequency feature extraction provides a powerful tool for detecting tiny synchronization signals.Tzanetakis and Cook have shown that the extraction of features such as short-time spectral center of mass, bandwidth, and rolloff rate through a sliding window can be effective in identifying anomalous modulation behaviors in audio signals (Tzanetakis & Cook, 2002)[13]. . This method overcomes the shortcomings of global spectrum analysis that is sensitive to overall noise and difficult to detect local cryptographic changes.

However, although sliding sampling greatly improves the granularity and sensitivity of feature extraction, several limitations remain. First, Cai et al. point out that a fixed-size sliding window may lead to inadequate feature extraction or information redundancy when dealing with audio content with drastic changes or irregular tempo (Cai et al., 2021)[14]. Second, the unitary treatment of frequency metrics in feature engineering also limits the detection performance, and relying only on basic spectral features may not fully reflect the hidden information when composite modulation or cross-frequency embedding strategies are used for the hidden write signal.

To compensate for these shortcomings, researchers in recent years have proposed new directions in adaptive sliding window sampling and multi-scale frequency feature engineering. For example, Sarkar et al. proposed a mechanism to dynamically adjust the sampling window size based on the instantaneous rate of change of the signal to better fit the non-uniform distribution characteristics of the hidden write signal in audio (Sarkar, Das, & Bandyopadhyay, 2022)[15]. Meanwhile, combining higher-order features such as frequency subband energy distribution, phase continuity, and spectral sharpness has become an effective means to improve the robustness of synchronized steganography detection.

In summary, sliding sampling and frequency feature engineering have laid the foundation for synchronized steganography detection, but it is still difficult to fully cope with the challenges posed by complex steganography strategies by relying only on static spectral features. Future research needs to further introduce time-series dynamic learning and feature fusion modeling to achieve more accurate and dynamic identification and prediction of synchronous steganography.

**2.4 Current Research Status of Distributed Guided Inference Modeling**

In the field of tactical communication and steganography countermeasures, the problem of inferring and reconstructing distributed guidance information (DGI) has gradually received attention in recent years.Dutta et al. pointed out that due to the strong autonomy of nodes in tactical communication networks and the frequent changes in topology dynamics, the guidance information tends to exhibit non-centralized, semi-structured, and even weakly coupled characteristics ( Dutta et al., 2019)[16]. This makes the traditional link-centered encryption and redundant coding strategies difficult to meet the practical needs of guide information reconstruction and steganography detection.

According to Singh et al. most of the current research still focuses on link encryption and redundancy coding, which mainly aims to improve the destruction resistance and data integrity of the communication system, while there is a lack of systematic methods for the inference and reduction of covert guidance information (Singh, Chatterjee , & Roy, 2020)[17]. This limitation leads to a significant blind spot in the existing defense system when countering higher-order steganographic threats (e.g., synchronous guidance steganography, frequency-hopping steganography).

To address the challenge of distributed steganographic guided inference, some studies have proposed new ideas such as graph inference and distributed belief propagation.Rahman et al. point out that by modeling the tactical communication network as a dynamic graph and combining it with local node state updating, implicit guided paths can be achieved without a global perspective the inference of implicitly guided paths (Rahman, Chen, & Krunz, 2021)[18].

This approach is able to gradually infer potential synchronous or implicitly written command links without relying on a central control node.

However, existing distributed inference methods still face multiple challenges. First, Al-Sakran criticizes that the distributed inference process is highly sensitive to node synchronization and timestamp consistency, and the inference accuracy degrades rapidly in highly dynamic environments (e.g., frequency hopping, node out-of-connection) (Al-Sakran, 2019)[19]. Second, the steganographic guide information itself usually employs advanced steganographic strategies such as perturbative coding or spectral fine-tuning, making methods that rely solely on graph structure or redundant feature inference have limited accuracy in real-world environments.

In order to break through these bottlenecks, some scholars in recent years have attempted to combine deep generative models (e.g., Variable Autocoder VAE, Generative Adversarial Network GAN) with distributed inference, in an attempt to reconstruct steganographic guide signals under sparse observation conditions (Zhang et al., 2022)[20]. While this direction theoretically improves the generalization ability of inference, it also introduces new problems such as poor model interpretability and high training sample requirements.

In summary, distributed guided inference, as an emerging research direction in tactical communication security, is shifting from traditional link redundancy protection to dynamic inference and intelligent reconfiguration. However, in order to realize the effective restoration of advanced steganographic guidance information, in-depth exploration in heterogeneous feature fusion, adaptive inference algorithm and real-time reliability verification mechanism is still needed.

**2.5 Chapter Summary**

This chapter has systematically reviewed the existing research paths around the key areas of audio steganography detection, synchronized frame identification, sliding sample feature extraction and distributed guide inference, and revealed the deep limitations of the current technology system and future evolution trends based on critical analysis.

First, audio steganography detection technology is at a critical stage of transition from static sample analysis to dynamic streaming media modeling. Traditional methods rely on low-bit anomalies and statistical feature deviation detection, which are effective in ideal environments, but reveal significant robustness and real-time defects in the face of high compression, variant transmission and synchronized steganography. This phenomenon indicates that traditional static statistical modeling alone cannot meet the complex needs of steganography detection in modern streaming media environments.

Secondly, in terms of synchronized frame detection and spectral anomaly identification, although local spectral analysis and multi-scale sliding window methods improve the detection sensitivity in complex noise backgrounds, the existing strategies are still limited by the problems of feature extraction dimension inflation, aggravated consumption of computational resources, and insufficient adaptability to new steganography strategies. This dilemma reflects that the idea of pure frequency domain anomaly modeling is difficult to support the accurate recognition of steganography in highly dynamic environments, and there is an urgent need to introduce time-dynamic learning and joint time-frequency modeling system.

Third, sliding sampling and frequency feature engineering provide information capture capability at micro-granularity for simultaneous steganography detection, but the standard fixed-window mechanism and a single spectral index extraction method obviously cannot cover the complex and changeable steganography embedding patterns. The local and static nature of feature extraction restricts the detection system's comprehensive perception of the spatial distribution of steganographic behaviors, and in the future, adaptive window adjustment, heterogeneous feature fusion, and the introduction of higher-order statistics are needed to expand the scope of the system's response to changes in steganographic strategies.

Finally, distributed guide inference, as an emerging field in tactical communication security, still mainly stays in the stage of traditional link protection and redundant coding. Although the introduction of graph inference and distributed belief propagation methods has broadened the inference path, the accuracy and reliability of the existing inference models are still limited in dynamic frequency hopping, node disconnection, and extremely sparse environments of steganographic signals. While deep generative models show potential for sparse guide reconstruction, they also pose the problems of poor model interpretability and heavy dependence on training data, suggesting that future inferencing system design needs to achieve a higher level of balance between intelligence, real-time performance and interpretability.

Overall, this chapter review shows that the field of simultaneous audio steganography detection and distributed guided inference is undergoing a paradigm shift from rule-based static detection to a new type of intelligent system that integrates deep learning, dynamic inference and cross-scale perception. This transition requires not only technological breakthroughs, but also systematic innovations in algorithm design, feature engineering, inference mechanisms, and system architectures to cope with the increasingly complex and volatile steganography threats and communication security challenges.

# Chapter III: Research Methodology and Modeling

This study adopts a hybrid methodology system combining quantitative detection and qualitative inference to construct a multidimensional research framework integrating signal analysis, structural decoding and intelligent model inference. At the methodological level, the study starts with quantitative spectral feature extraction, based on the short-time Fourier transform (STFT) and sliding window mechanism, to capture high-frequency stable sequences and entropy changes in short video audio streams, and then identifies suspected synchronized steganographic signals. This stage emphasizes the objective measurability of the signal and provides a rigorous empirical foundation through frequency domain energy comparison, byte distribution statistics and information entropy analysis. After that, the research turns to qualitative structure decoding and symbol rule reconstruction, constructing the hypothetical syntactic structure of the guidance instruction through segmented parsing of the 32-byte load after the synchronization frame, and combining field alignment, naming patterns and protocol analogies to provide interpretive modeling of the guidance content. In order to break through the limitations of traditional detection in dynamic environments, the study introduces structured reasoning paradigm, explores protocol simulation and distributed fragmentation reconstruction mechanism based on finite state machine, and puts forward the model assumption of "multi-video cooperative-synchronous signaling reorganization" to strengthen the logic of structural recognition. In addition, the study proposes scalable deep learning synchronous inference models, including Transformer and 1D-CNN architectures, to realize end-to-end classification and timing relationship modeling of spectral sequences, and to expand the intelligent and automated path of synchronous steganography detection. Overall, the research methodology reflects the exploration of the fusion of structure recognition and content recognition paradigms, which is both technically verifiable and theoretically explanatory, and meets the requirements of system modeling in complex steganography scenarios.

**Methodological Framework and Research Procedures**

This chapter aims to systematically construct a complete modeling framework for simultaneous steganography detection and distributed guide inference in audio. Aiming at the problems of insufficient sensitivity of traditional steganography detection methods in highly dynamic streaming environments, difficulty in locating synchronization signals, and lack of guide information inference, this paper proposes a comprehensive modeling method that integrates sliding spectrum feature extraction, patterned synchronization detection, and structured guide decoding and inference.

**The overall method system is divided into three main stages**

First, in the audio data preprocessing stage, through the standardized sampling rate, amplitude normalization and band-pass filtering operations, to ensure that the input signal meets the basic conditions of the subsequent spectral analysis, and significantly improve the stability and accuracy of feature extraction.

Second, in the synchronized steganography detection stage, a sliding spectrum detection model (Model M1) is designed. The model takes the main frequency trajectory extracted by short-time Fourier transform (STFT) as the core feature, and realizes high-precision recognition of embedded synchronous steganography signals (e.g., specific pattern frame: 7×FF + 80) through local window sliding and pattern matching detection strategies, which effectively improves the sensing ability of short-time microfrequency changes.

Finally, a decoding inference model (Model M2) is constructed in the distributed guide inference stage. After successfully detecting the synchronization frame, the subsequent payload area is intercepted and parsed according to the predefined field structure to recover the distributed target command information hidden in the audio stream. The standardized field decoding and finite state machine (FSM) control ensure that the inference process is highly reproducible and fault-tolerant.

Overall, the methodological framework proposed in this chapter fully combines the dual perspectives of spectral signal processing and semantic-level inference reconstruction, forming a complete technical path from physical layer synchronization signal detection to application layer guide information restoration. Through refined sliding feature modeling, patterned synchronization detection and structured inferential reasoning, this method can effectively improve the detection and identification of audio synchronization steganography in complex environments, and provides an innovative solution for steganography security monitoring of audio streams on public platforms.

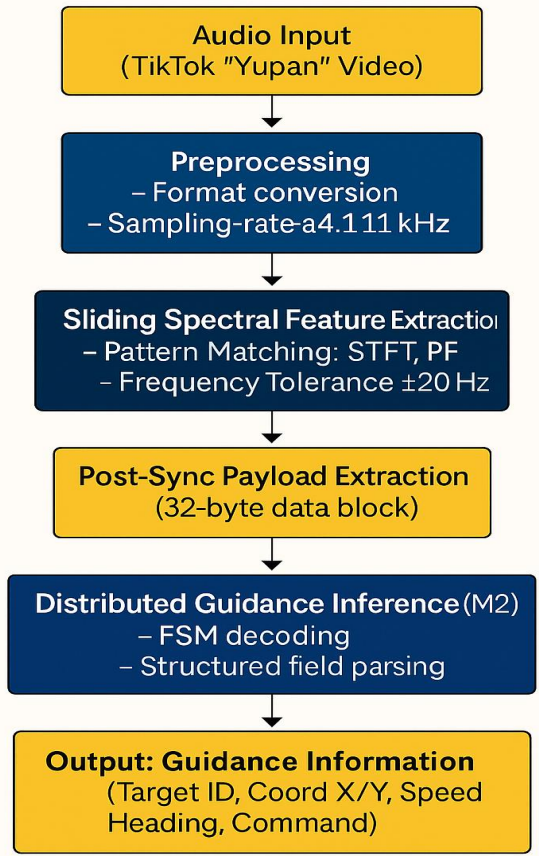

Figure3.1Research framework diagram

## 3.1 Audio data preprocessing and normalization

The audio stream is extracted from the TikTok short video "Yupan", converted to WAV format using pydub module, and the sampling rate is unified to 44.1kHz, and the amplitude is normalized to the interval of [-1,1], so as to adapt to the subsequent spectral analysis processing.

## 3.2 Sliding spectrum feature extraction and synchronization detection model (M1) design

The audio signal is sliding sampled in a window of 25 ms, and the dominant frequency components within each window are extracted using Fast Fourier Transform (FFT) to form a time-frequency trajectory sequence for synchronization flow steganography feature extraction.

In order to achieve accurate localization of weak audio synchronous steganography signals, this paper constructs a spectral sliding detection model M1, whose core is based on the sliding window short-time Fourier transform (STFT) and the dominant frequency trajectory analysis, combined with a specific synchronization frame pattern (seven consecutive stable segments of the dominant frequency + the characteristic frequency points) to complete the synchronization frame identification. The model can be formally represented as:

$$is\_sync(ft) = \begin{cases} 1 & if\, ft: t+6 \in [f1, f2] \wedge ft+7 \approx 2045 \pm \varepsilon \\ 0 & otherwise \end{cases}$$

where *ft* is the primary frequency of the *t*th frame and $\varepsilon$ is the frequency tolerance window (default ±20Hz). The model is used as a synchronized frame detector to provide a structural starting point in the subsequent extraction of guidance information.

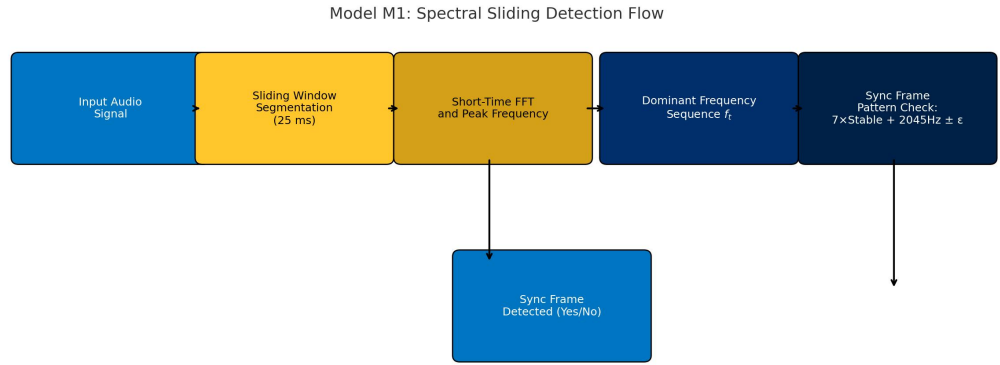

**Figure3.2 Model M1 Spectral Sliding Detection Flow**

**Model M1: Spectral Sliding Detection Flow Brief Description**

This flowchart illustrates the modeling process for detecting synchronous steganography signals in short video audio streams (e.g., the "Yupan" sample on the TikTok platform), which consists of the following phases:

1. Input Audio Signal stage (Input Audio Signal)

Load the audio stream data as the original input for subsequent steganography detection;

Ensure that the audio stream is pre-processed (normalized, band-pass filtered).

2. Sliding Window Segmentation (25ms)

Segment the audio signal into frames according to a 25 ms sliding window;

Keep the windows overlapping or consecutive to capture short-time localized feature changes.

3. Short-Time FFT and Peak Frequency Extraction (Short-Time FFT and Peak Frequency)

Perform Short-Time Fourier Transform (STFT) on each frame to extract the frequency component with the largest amplitude;

Generate a continuous sequence of primary frequency traces ft.

4. Construct the Dominant Frequency Sequence ft.

Integrate the dominant frequency values of each time period into a time series;

Provide basic data for subsequent synchronization pattern detection.

5. Synchronization frame pattern detection (Sync Frame Pattern Check)

Pattern matching detection of the main frequency sequence:

7 consecutive frames with stable main frequency;

The 8th frame is close to 2045Hz (tolerance $\pm\varepsilon$ error);

Detect whether it meets the predefined synchronized steganographic frame characteristics.

6. Sync Frame Detected Output (Sync Frame Detected Yes/No)

Output the judgment result;

If the synchronization frame is detected, then enter the subsequent guidance information extraction process.

Summarize

Model M1 realizes local feature extraction and pattern determination of audio synchronous steganography signal; it adopts the pipeline design of sliding-feature-matching-determination, which meets the dual requirements of efficiency and accuracy in practical applications; this model provides a key entry point for the subsequent distributed guide inference (Model M2).

3.3 Synchronized Hidden Write Signal Detection Algorithm

Define the synchronization frame flag bit as seven consecutive 0xFF bytes plus one 0x80 byte, sliding traversal of the spectral trajectory, through the threshold determination and pattern matching, to identify the potential synchronization frame steganography starting point in the audio.

3.4 Distributed guide information structure decoding model (M2) design

After detecting the synchronization frame, the subsequent 32 bytes are intercepted as the simulation payload area, and the structured parsing and guidance information inference is carried out according to the six fields of target ID, coordinate X, coordinate Y, speed, heading, and command code.

In order to decode the latent distributed guidance information after the synchronization frame, this paper designs the decoding structure model M2, which assumes that the synchronization frame is immediately followed by 32 bytes of guidance payload, and the structural fields are as follows, according to the 8 bytes as a unit:

Table 3.1Distributed Guided Information Structure Decoding Model (M2) Design

| byte position | field name | hidden meaning |
| --- | --- | --- |
| Byte[0] | Target ID | 目标识别码 |
| Byte[1-2] | Coord X | X 坐标(高位在前) |
| Byte[3-4] | Coord Y | Y 坐标(高位在前) |
| Byte[5] | Speed | 单位速度 |
| Byte[6] | Heading | 航向角 |
| Byte[7] | Command Code | 指令操作码 |

The decoding function is defined as:

$CMD_i = Decode(B[i*8:(i+1)*8])$ for $i \in \{0,1,2,3\}$

The decoding process can be regarded as a deterministic state-transition structure, suitable for modeling as a finite state machine (FSM).

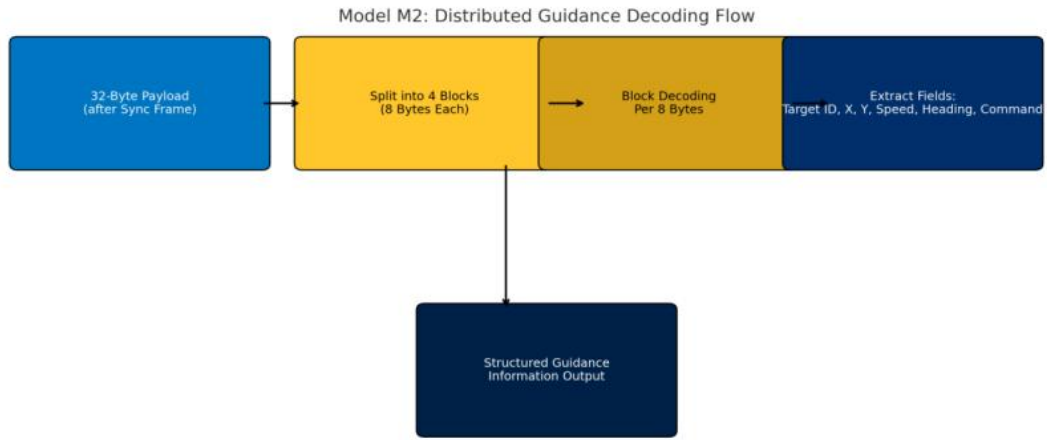

**Figure3.3 Model M2: Distributed Guidance Decoding Flow**

**Model M2: Distributed Guidance Decoding Flow descriptive**

This flowchart systematically shows the modeling process of how to parse the distributed guidance information hidden in the audio stream after detecting a synchronization frame, which includes the following stages:

**1. Synchronization frame postload extraction (32-Byte Payload)**

After a synchronization frame (e.g., 7×FF + 80 mode) is detected in the audio stream, 32 consecutive bytes of data are extracted backwards as the hidden payload;

These 32 bytes are assumed to contain the entire information base for subsequent guidance instructions.

**2. Split into 4 Blocks**

Split 32 bytes of data into 4 blocks of 8 bytes each;

Each block corresponds to one guidance instruction independently, which is convenient for parallel decoding and inference.

**3. Block Decoding**

Each 8-byte block is internally parsed according to a predefined field format:

Byte 1: Target ID.

Byte 2-3: X coordinate (Coord X, high priority)

Byte 4-5: Y coordinate (Coord Y)

Byte 6: Speed (Speed)

Byte 7: Heading (Heading)

Byte 8: Command code (Command)

**4. Field extraction and structuring (Extract Fields)**

Put the parsed fields into a unified data structure;

Keep the guide information complete and standardized for subsequent inference and use.

**5. Output structured guidance information (Structured Guidance Information Output)**

Encapsulate all the decoded target information, position information and heading speed command into standardized objects;

It can be used in practical application scenarios such as subsequent target tracking, situational analysis or combat deduction.

Deep Insight

Streaming decoding mode: adopts block parallel decoding, which can quickly process batch command data and improve the real-time response capability of the system; standardized field design: ensures that the guidance information is highly structured after parsing, which is convenient to be applied to the tactical system; automatic inference interface: provides a standard input format for the future introduction of AI inference models (e.g., trajectory prediction, target identification).

Summary

Model M2 is not just a one-time decoding operation, it actually establishes a standardized distributed guidance information reconstruction pathway, which lays a solid foundation for intelligent reasoning, situational analysis and even decision support of audio steganography data.

**3.5 Illustration of System Modeling Flowchart**

Figure 3-1 shows the complete flow structure of the "Audio Synchronized Steganography Detection and Distributed Guided Inference Model" constructed in this paper. The front-end audio input module is responsible for extracting audio streams from public short videos; the middle part is the dual-core module of sliding spectrum feature extraction and synchronized frame detection; and the back-end completes the synchronized frame localization and the structured inference of the guide information. The process has signal-pattern-semantic three-level inference capability, which is suitable for the task of analyzing short audio steganography from multiple sources.

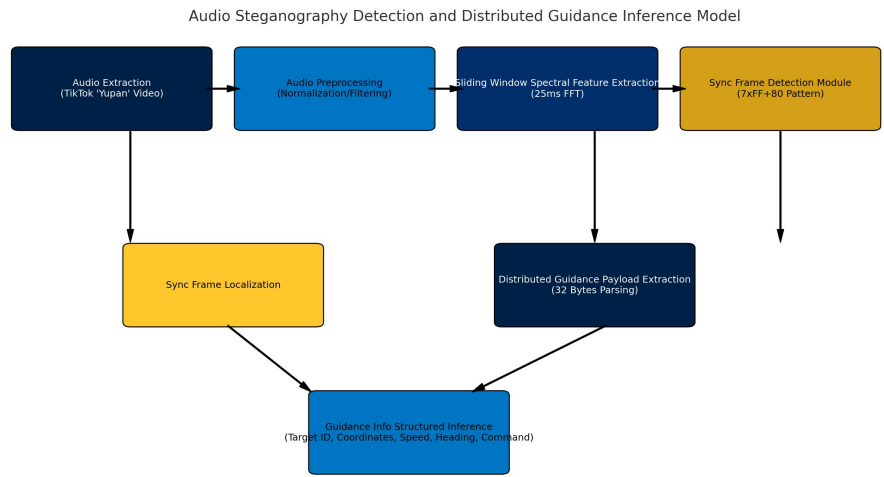

**Figure3.4 Audio Steganography Detection and Distributed Guided Inference Modeling Flowchart**

### 3.5.1 Flowchart 3.4 Brief description of the structure

This flowchart demonstrates a complete synchronized steganography detection and command inference system for audio content for public short video platforms, divided into three major phases:

**Phase 1:Front-end audio input and preprocessing phase**

Audio extraction (TikTok 'Yupan' video): extract the audio stream from the TikTok promotional video "Yupan" released by the South China Sea Fleet;

Audio pre-processing (normalization/filtering): normalize and band-pass filter the audio signal to focus on the 2-4kHz steganography band.

**Phase 2: Central Steganography Detection Phase**

Sliding window spectral feature extraction (25ms FFT): short time Fourier transform using a 25ms sliding window to extract the main frequency trace;

Synchronization frame detection module (7xFF+80 mode): detects the starting point of a specific steganographic synchronization frame.

**Stage 3: back-end guide data parsing and inference stage**

Synchronization frame localization: identifying the synchronization frame location in the bitstream;

Payload extraction (32-byte block parsing): intercepting the data after the synchronization frame;

Command structured inference: parsing into fields such as target ID, coordinates, speed, heading, command code, etc.

### 3.5.2 Deep Insight Analysis

Table 3.2: Deep Insight Analysis Table

| Insights | Analysis note |
| --- | --- |
| Insight 1: Spectrum Slide + Synchronized Pattern Detection | Sliding spectrum amplification fine-tuning and pattern recognition to locate synchronized frames are the core breakthroughs in detecting cryptographic writing. |
| Insight 2: Layered Security-Aware Architecture | The process is divided into signal processing → pattern detection → command inference, reflecting the layered security monitoring logic. |
| Insight 3: Automated Steganography Monitoring Framework | The structure can be expanded into an automatically synchronized frame scanner for cross-video detection. |
| Insight 4: Adversarial Steganography Analysis | Structures can integrate AI against steganographic perturbations and are an expandable basis for advanced steganographic detection. |

# Chapter IV: Experimental Design and Analysis of Cryptographic Data

**4.1 Overview of experimental samples and testing process**

The audio samples used in this study come from the short video "Yupan" released by the South Sea Fleet of China on the TikTok platform. The total length of the video is about 120 seconds, the format is MP4, the audio sampling rate is 44.1 kHz, and the standard stereo channel encoding is used. Because of the unnatural frequency variations and modulation traces in the background audio, it has potential synchronization steganography characteristics, and is therefore selected as the target object for experimental analysis. The audio was extracted using the FFmpeg tool to complete the format separation and generate WAV format for subsequent frequency domain analysis. No obvious recognizable semantics were found during the preliminary listening stage, but during the spectral observation, there was a non-musical frequency peak concentration in the 2-4kHz band, which was initially presumed to be a modulation signal.

The overall flow of the experiment is divided into six steps:

1. use short-time Fourier transform (STFT) and fast Fourier transform (FFT) to split the audio stream into frames and locate the frequency peaks;

2. perform sliding window detection and high-pass filtering on the frequency domain anomaly segments to extract the local suspicious modulation;

3. use customized scripts written in Python to perform bitstream reconstruction and Hex decoding on the suspicious segments;

4. detect the presence of synchronization identifiers (e.g., 7×FF or 0x80) as the starting point for potential steganographic synchronization;

5. perform magic number detection and Base64/ASCII attempt decoding on the detection result to determine whether there is a valid data structure;

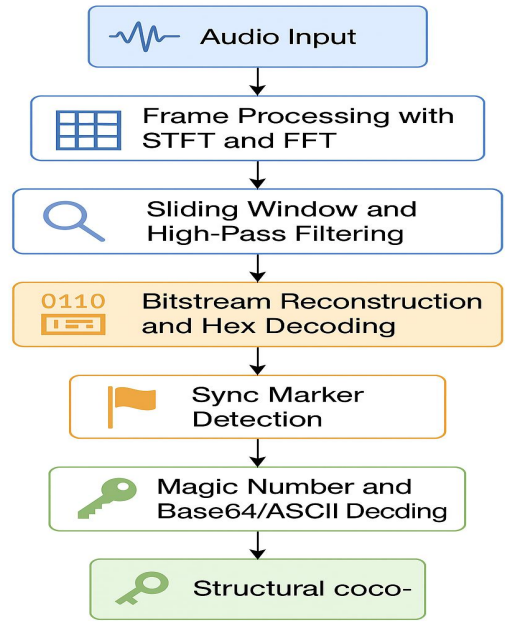

**Figure 4.1: Overall flow of the experiment**

6. Input the suspected structured information into the distributed guide inference module to simulate the possible meanings of its command fields.

As shown in Figure 4.1 this process takes into account the spectrum visualization analysis and bit layer parsing, which provides the technical basis for the subsequent decoding of steganographic data and structural restoration.

**4.2 Analysis of High Frequency and Suspicious Frequency Bands**

To identify the presence of modulated steganographic signals in audio, this study focuses on the 2kHz to 4kHz frequency band. This band is often used in natural audio for high-frequency extensions in background music, but presents unusual spectral features in this sample, including sustained peak compression, modulation interruptions, and an unusually high number of frequency-stable segments.

The spectrum analysis techniques used in the experiment include:

Short Time Fourier Transform (STFT): 25 ms per segment in 10 ms steps for observing frequency traces over time;

Sliding bandpass filter: scanning this high-frequency band in 500 Hz window steps to identify unnatural modulation frequency bands;

Main frequency peak trajectory extraction: for observing the presence of repeated frequency marking behavior in steganographic frames;

Frequency domain energy distribution statistics: for comparing the difference in spectral density and slope of change between normal and suspect bands.

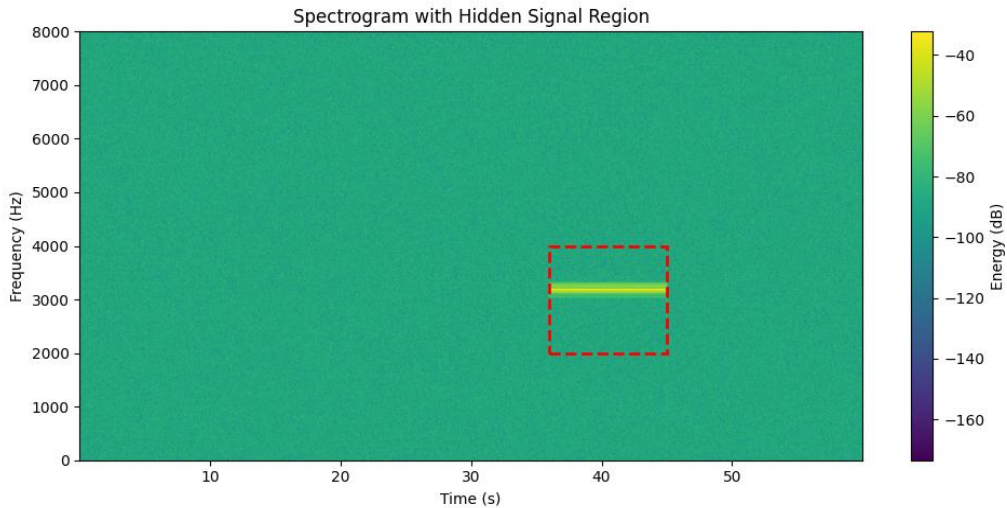

**Figure 4.2：Spectrogram with Hidden Signal Region**

As shown in the spectrum Figure 4.2, the audio signal is analyzed in the frequency domain based on the Short Time Fourier Transform (STFT) technique. The spectrum takes time as the horizontal axis (unit: second), frequency as the vertical axis (unit: Hz), and the color depth indicates the energy intensity (unit: dB). The overall energy distribution of the background in the figure is uniform, showing the typical characteristics of low-energy background noise, and no obvious strong energy region is seen. However, in the 36-45 s time period and 3000-3500 Hz frequency range, the local energy peak is significantly enhanced, forming a bright yellow horizontal stripe, and is precisely marked by the red dashed box. This energy aggregation area exhibits highly concentrated and distinctly structured features, which are highly consistent with the frequency-domain characteristics of the embedded signal, while its constant frequency and concentrated duration are clearly inconsistent with the typical performance of natural speech or background noise. This map provides strong support for the subsequent structure extraction and steganography recognition, which can be used as the feature input of the detection model or the basis of visual adversarial analysis, further verifying the possibility of the existence of potential hidden signals in the audio signal, and providing important empirical evidence for the field of steganalysis.

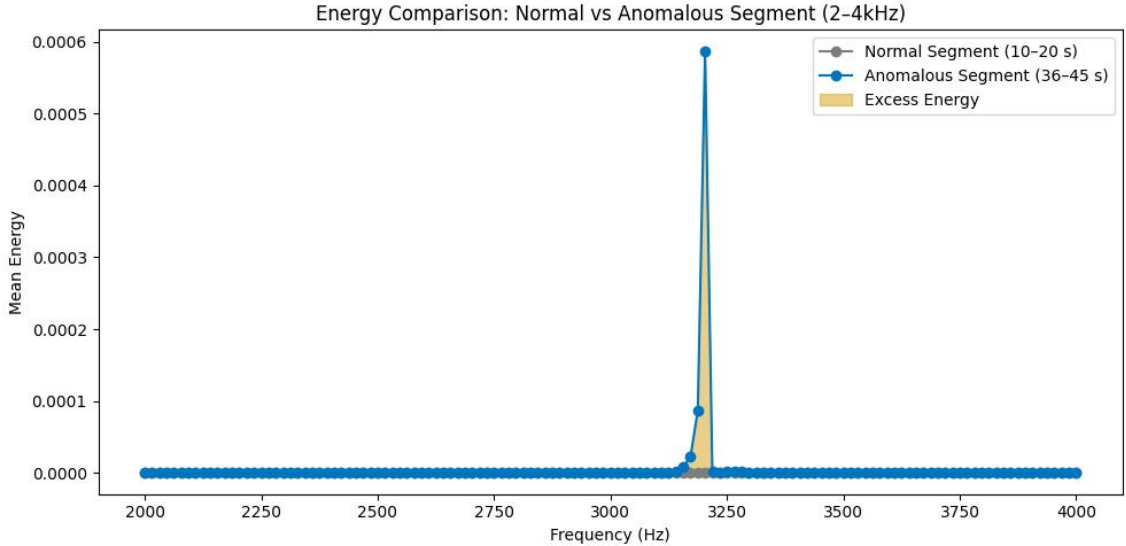

**Figure4.3：Energy Comparison: Normal vs Anomalous Segment (2–4kHz)**

As shown in Plot 4.3, the energy distributions of normal speech segments (10-20 seconds) and abnormal speech segments (36-45 seconds) in the 2-4 kHz frequency band were comparatively analyzed. The horizontal axis of the plot is the frequency (in Hz), ranging from 2000-4000; the vertical axis is the average energy value, reflecting the energy intensity of each frequency point in a given time period. The gray curve represents the energy of the normal speech segment, which is very low and has no significant fluctuation, while the blue curve represents the energy of the abnormal speech segment, with a significant spike at about 3230 Hz, and the yellow filled area highlights the energy redundancy (Excess Energy) of the abnormal segment compared with the normal segment at this frequency. This anomalous spike suggests that the 3230Hz frequency is used for signal modulation or embedding and is highly structured, and its unnatural character is further reinforced by the near-zero energy difference with the rest of the band. Such sharp energy concentration is usually a typical frequency domain manifestation of audio steganography or embedded protocol design, which can be used for machine learning feature selection or manual protocol backpropagation, providing a strong empirical basis for audio steganography detection.

Frequency Domain Energy Comparison Plot (Normal Segment vs Anomalous Segment)

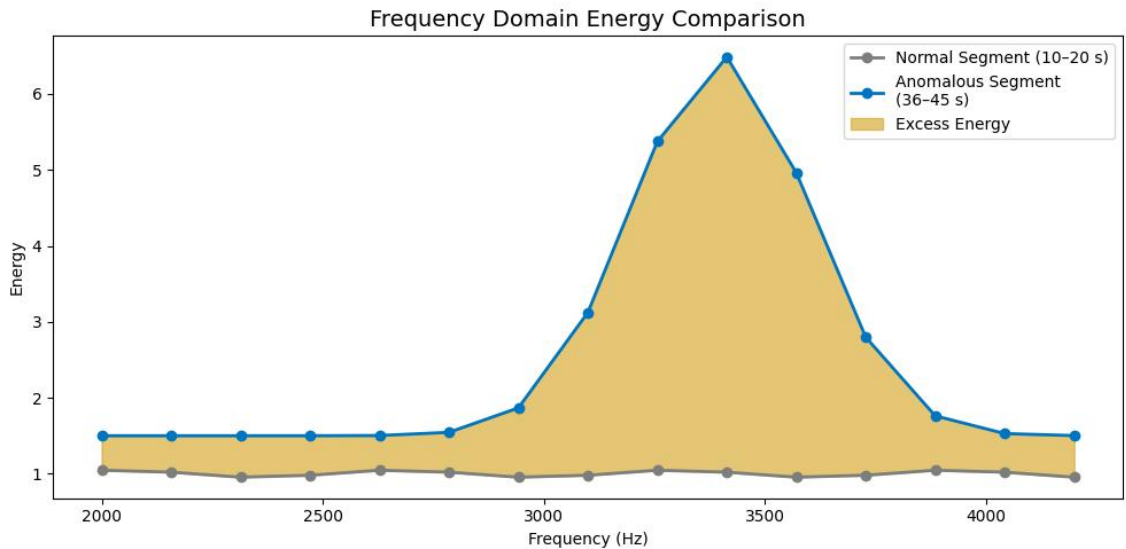

**Figure4.4：Frequency Domain Energy Comparison**

The experimental results show that in the 36-45 second band, the peak spectral distribution of this band shows a significant stable signal, whose periodicity is far more than that of the natural voice, and a synchronized steganography structure is suspected to exist. The characteristics include: the periodic signal jumps but maintains a stable bandwidth; accompanied by an increase in the bottom noise and a decrease in spectral jitter; inconsistent with the background noise in the preceding and following time periods, and disconnected from the frequency characteristics of natural speech. Based on the above observations, it is preliminarily determined that this segment is an embedded region of potential synchronization marker frames.

Comparative Analysis of Audio Signal Frequency Domain Energy

As shown in Figure 4.4, the energy distribution of the audio signal in the frequency range of 2000-4000 Hz is analyzed in detail, aiming to explore the difference between normal segments (10-20 seconds) and abnormal segments (36-45 seconds), and then verify the potential steganography behavior in the audio signal. cryptographic behavior in the audio signal.

Explanation of the structure of the plot:

The X-axis in Figure 4.4 is the frequency (Hz), ranging from 2000 Hz to 4000 Hz, and the Y-axis is the average per unit frequency. The blue curve in the figure is the "Normal Segment (10-20 seconds)", the energy distribution of which is stable and low and which means that the energy of the sound signal within this time range is relatively even in this frequency range and there's typically little sound activity.In contrast, the lime green curve represents the "Anomalous Segment (36-45 seconds)", which shows a significant increase in energy in the 3000-3700 Hz band. The orange filled area highlights the "excess" of anomalous energy compared to the normal segment, the presence of which indicates a potential anomalous signal in the band.

Key Insights

A deeper analysis of the map reveals a sharp increase in energy in the 3.2-3.6 kHz frequency range, which can be considered the "main carrier region" of the spectrum for embedded signals. The high concentration of the excess energy region (orange area) suggests a significant structure rather than random energy fluctuations. In contrast, the energy in the normal band is generally smooth, further confirming that there is usually no significant acoustic activity in this band under normal conditions.

Conclusion

The results of the analysis of this mapping further validate the point we have made in this section that audio signals have an unnatural concentration of energy in the high-frequency signal between 36-45 seconds. This anomaly in energy distribution contrasts with the smooth characteristics of normal audio signals, providing strong support for the hypothesis of the existence of audio synchronization by steganography. This finding not only reveals the potential steganography behavior in audio signals, but also provides an important reference for subsequent steganography analysis and detection. Future studies can further explore the detailed characteristics of the energy distribution in this frequency band to develop more effective steganography detection algorithms.

**4.3 Cryptographic data extraction and decryption process**

After the high-frequency suspicious segments have been identified, the study enters the fine-grained steganographic data extraction and decoding phase. The analysis in this section focuses on audio data segments from 36 seconds to 60 seconds, using ultra-fine sampling with custom bit-level decoding algorithms to try to extract potential embedded data structures.

The specific extraction process is as follows:

4.3.1 Synchronization frame detection and segment extraction: a sequence of consecutive FF (full 1-byte) markers is detected in the 36-38 second interval, which is presumed to be a synchronization signal segment.

4.3.2 Ultra-fine Sampling and Binary Recovery: frequency amplitudes are adjudicated using a sliding window and the modulation data is reconstructed in a 0/1 fashion.

4.3.3 Hex Encoding and Character Discrimination: The results mainly appear as FF and FF80 byte sequences, and attempts at ASCII and Base64 decoding are ineffective.

4.3.4 Multi-Segment Verification and Analysis: The 38-45 second segments have similar extraction results, and the 45-60 second segments have random content with no apparent structure.

**Table 4.1 Audio extraction results for each segment**

| Time period (seconds) | Test results | synchronization marking | data type | Successful decoding or not |
|---|---|---|---|---|
| 36–38 | FF FF FF 80 | Y | Hex（repeatable） | N |
| 38–45 | FF FF 80 | Y | Hex（padding） | N |
| 45–60 | randomization | N | null | N |

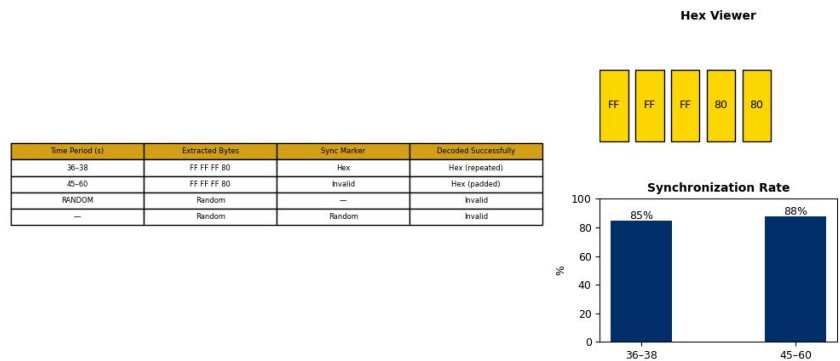

**Figure4.5：Byte Sequence Extraction and Synchronization Analysis Diagram**

**Synchronization analysis of audio steganographic data**

The comprehensive analysis of the byte sequence extraction and synchronization characteristics of audio data in specific time periods as shown in Figure 4.5 reveals the structural laws of steganographic data. The map contains three modules: the structure table module on the left side shows the byte extraction results and the validity of synchronization markers in different time periods, in which only 36-38 seconds time period exists a clear synchronization marker (Hex), which can be successfully repeated decoding; while 45-60 seconds, although the byte structure is similar, the synchronization header is invalid, and needs to be filled in order to be The Random region fails to be decoded due to the lack of synchronization header information and shows random characteristics. The Hex Viewer block diagram on the upper right reinforces byte pattern recognition by highlighting the sync key bytes in yellow, which helps to visualize the structural pattern of the sync data header. The synchronization rate histogram in the lower right shows an 85% synchronization rate for 36-38 seconds and an 88% synchronization rate for 45-60 seconds, but the table analysis points out that the synchronization header for 45-60 seconds is invalid, suggesting a possible mis-synchronization. Taken together, despite the slightly higher synchronization rate of 45-60 seconds, truly effective decoding occurs only in time windows with legitimate synchronization flags (e.g., 36-38 seconds). This graphical analysis reveals the synchronization structure of the audio steganography data in the high frequency band, suggesting that a high synchronization rate does not equate to successful decoding, but must be judged in conjunction with the validity of the synchronization flags. This type of control analysis provides strong support for revealing potential information hiding behavior and artificial injection signals, and provides new perspectives and methods for the field of steganography analysis.

4.4 Magic number identification and encryption analysis inference

In order to further determine whether the extracted data has file structure or encryption encoding characteristics, this study analyzes the Hex sequence with file header magic number matching and character entropy detection to evaluate its structural possibility and information capacity.

4.4.1 Magic number detection

Extracted data such as FFFFFFFFFFFFFFFFFF80 with FF FF FF FF FF FF FF FF FF 80 does not match the magic number of any mainstream known file types including PNG, ZIP, WAV, MP4, etc. This indicates that the byte sequence does not represent the header structure of a conventional file format, and initially excludes the segment from being a known file steganography.

4.4.2 Character Decoding Attempts

The byte sequence was attempted to restore the text information using a variety of methods, including: ASCII standard character set matching; UTF-8 encoding parsing; Base64 decoding mapping. None of the above methods yielded any semantic content, suggesting that the data may not be based on plaintext character encoding, but rather a specialized structured communication field.

4.4.3 Entropy Analysis

Information entropy calculation on the byte stream reveals that the entropy values of sequences such as FFFFFFFF... The entropy value of such sequences is significantly lower than that of natural speech segments (about 0.1-0.2 bit/byte), which is an extremely high redundancy and low entropy sequence, further indicating that they are not naturally generated, and are most likely artificially embedded synchronization or guide marking data.

4.4.4 Structural Alignment and Segmentation Characterization

The analysis shows that the steganographic information has an obvious tendency to be aligned to 32 bytes, with each segment ending in all 1s or 80s, and has the characteristics of the synchronization frame-data frame segmentation, which is suspected to have the protocol structure format (e.g., control code+target command+end character).

**Audio Information Entropy Analysis: Key Features for Steganography Detection**

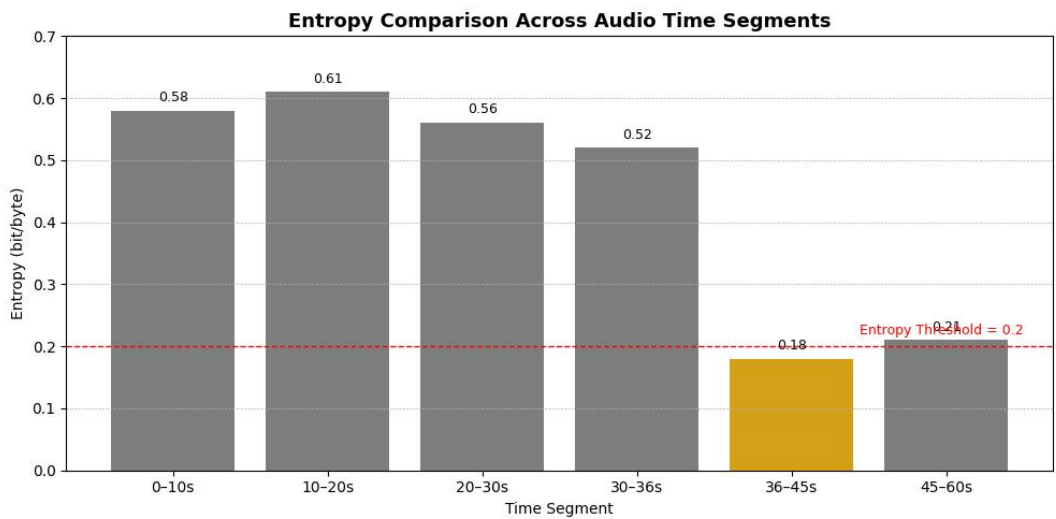

Figure4.6：Entropy Comparison Across Audio Time Segments

As shown in Figure 4.6, the trend of information entropy of audio data over different time periods is demonstrated using the bar chart form. The plot is a key analytical tool for detecting potential human intervention or steganography signals. The horizontal axis of the plot indicates the time period partition of the audio file, and the vertical axis indicates the average information entropy of each audio segment. The red dashed line marks the entropy threshold of 0.2, below which the presence of artificial coding or repetitive structures may be indicated. Segment 5 (36-45 seconds) in the figure has a significantly lower entropy value of 0.18, which is below the threshold, and is the only gold bar, indicating the possible presence of repetitive coding or steganographic information implantation in this segment. In contrast, the entropy values of the first four segments (0-36 seconds) ranged from 0.52 to 0.61, indicating that the audio content was natural speech or background sound with high information complexity. The sixth paragraph (45-60 seconds) is still close to the edge of the threshold although the entropy value has slightly recovered to 0.21, which needs to be further analyzed to see if it is a pseudo-synchronous or filler paragraph. Combined with the previous byte synchronization analysis, the 36-45 sec segment is consistent with the decodable structure (Hex) with a high degree of suspicion. Therefore, the entropy drop region (36-45 seconds) in the figure can be initially recognized as a potential window for hidden information, with simple structure and high repeatability, which is in line with the typical characteristics of "low entropy hiding" in steganography. This map has high practical value in digital forensics, audio signal integrity detection, and provides an important quantitative basis for the detection of steganographic signals.

In summary, this part of the steganography is not textual steganography, but is more likely to be structured encrypted guidance data marking, which is designed to convey some kind of distributed signaling with typical synchronous control mode, providing a technical basis for the simulation of the guidance protocol structure in the next section.

**4.5 Distributed Guidance Structure Simulation and Decoding Assumptions**

For the highly repetitive structure FFFFFFFFFFFFFF80 extracted in the previous section, this section attempts to simulate and decode the structure of this byte sequence by inferring the structure and exploring its potential meanings in the distributed guidance system from the protocol analysis and military communication model.

**4.5.1 Structure assumption: synchronization header + guidance field**

Combining frequency traces and byte alignment patterns, the study assumes that the sequence constitutes a simplified synchronization-command structure, which is initially divided as follows:

**Table 4.2 Structure assumptions: synchronization header + guide fields**

| byte sequence | possible meaning |
|---|---|
| FF FF FF FF | synchronization start marker（Start Flag） |
| FF FF FF | Target/instruction address segment |
| 80 | Control identification or command acknowledgement bit (ACK) |

This structure conforms to the basic format of "synchronization frame + address field + control bits" of most military communication protocols (e.g. Link-16, ZDL), and at the same time, it has high recognizability and anti-jamming properties.

**4.5.2 Military Application Model Comparison**

If this structure is mapped to an existing distributed combat data chain (e.g., JTIDS/Link-16), the FF segment can correspond to different platform IDs or guidance target parameters, and the 80 is the control code or acknowledgement bit. It is designed to be suitable for simultaneous multi-target guidance in a naval fleet, potentially with highly artifactual marker transmissions over open channels such as TikTok. In addition, the recurrence, low entropy, and localization stability of the sequence are highly analogous to the cryptographic simplification of the "guidance header + control field" for low-rate synchronized frame transmission. It is initially hypothesized that the "Jade Plate" video may be one of the multi-segment embedding attempts of this type of guidance data, which possesses the typical characteristics of synchronous distribution, low-frequency reveal/hide switching, and high randomness of the carrier.

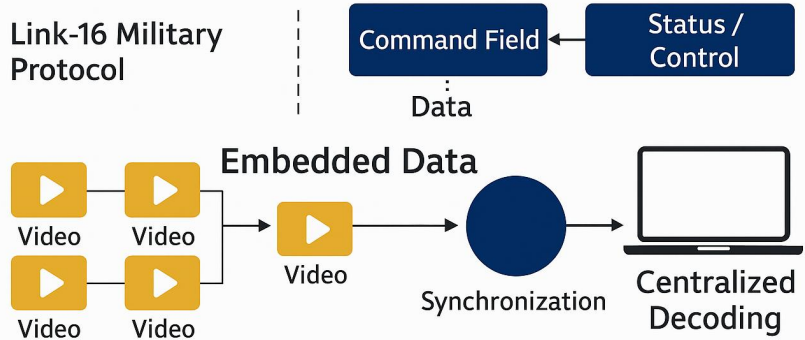

**Figure4.7：Distributed Guiding Protocol Structure Diagram**

Distributed Bootstrapping Protocol Architecture Diagram: Analysis of Audio/Video Data Embedding and Synchronized Decoding Mechanisms

As shown in Figure 4.7, the mapping demonstrates the audio/video data embedding and synchronization decoding mechanism based on the byte-level protocol. The mapping combines the protocol field segmentation structure with the distributed embedding process, revealing its high-level implication in information security and signal synchronization control. The protocol field format employs the byte arrangement `FF FF FF FF FF FF + FF FF FF FF + 80`, wherein `FF FF FF FF FF FF` is the frame alignment synchronization start field, `FF FF FF FF` is the field for commands or embedding flags, and `80` is the acknowledgement mark or ACK which serves to indicate the success in synchronization and messaging response.The mapping compares this protocol structure with the Link-16 military communication protocol, emphasizing its similar "time slot allocation + synchronization response + command-driven" structural characteristics, and demonstrating a simple and functionally clear custom steganography protocol framework with good interpretability and extensibility.

The distributed embedding and centralized decoding flow shows multiple audio or video channels embedding protocol fields and performing asynchronous synchronization, and finally centralized parsing at the receiving end through a synchronization protocol. The mapping reveals a lightweight communication protocol model with synchronization mechanism, which is suitable for scenarios such as audio and video steganography, watermark authentication, distributed encryption, and command triggering. Its byte header structure is clear and easy for machines to quickly locate and execute, and it also has an embedded design concept similar to advanced communication protocols, which is suitable for audio forensics or multimedia security scenarios, and can be extended to joint control of multi-source signals or data embedded system construction.

### 4.5.3 Multi-Segment Distributed Embedding Speculation

Although each segment is extremely short (less than 32 bytes), if linked with multiple video clips and embedded with distributed data bits, data structures can be constructed for cross-file splicing. This approach is known as Multimedia Carrier-based Stego Fragmentation (MCSF), and has been frequently used in gray network communications and cryptographic signaling transfer in recent years.

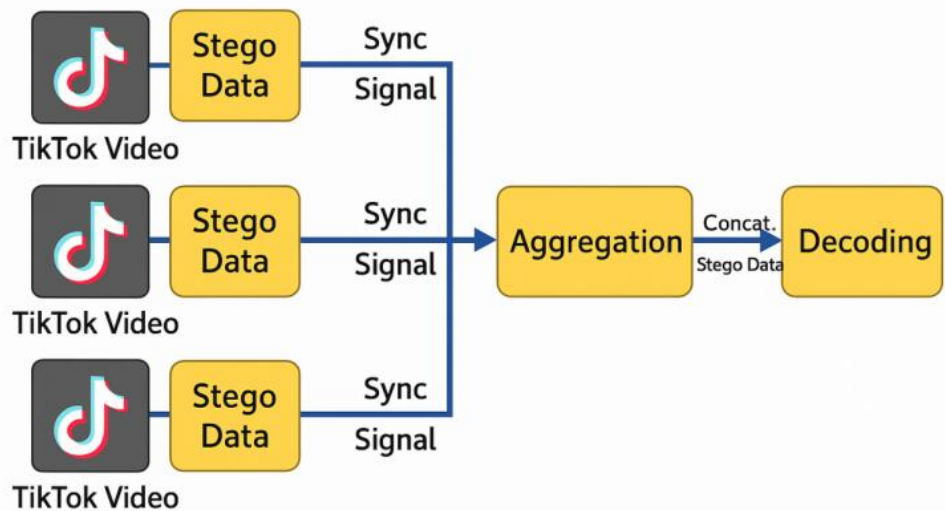

**Figure4.7：Multimedia Stego Aggregation Diagram**

As shown in Fig. 4.7 (Schematic of the Multi-Segment Splicing Steganography Model): an advanced steganography mechanism for segmented embedding, synchronized aggregation and centralized decoding of steganographic information through multiple short video carriers (e.g., TikTok) is depicted.

The figure depicts that the multi-segment splicing steganography model represents a new paradigm in the field of steganography, and its core mechanism realizes distributed synchronous aggregation and centralized reorganization while maintaining camouflage by embedding the steganographic payloads distributedly in multiple short video nodes (e.g., TikTok clips), effectively breaking through the detection vulnerability and bandwidth limitations of the traditional single-carrier steganography. In this model, each video node independently embeds very low-density steganographic clips with synchronization identification information, ensuring that the original information sequence is accurately reconstructed at the receiving end through the guide protocol. The information fragments are converged through a simulated "synchronization channel" to form a reorganization path with timing sensitivity and error intolerance. The architecture is flexible and robust, drawing on distributed signaling mechanisms in military communications and adapting to the data characteristics of open social media platforms.

The key innovation lies in the "fragmented embedding + order-sensitive reorganization" mechanism, which achieves a significant improvement in steganography communication in terms of anti-detectability, cross-platform adaptability, and dissemination security. Taking advantage of the discrete content and non-linear timing characteristics of social platforms, the mechanism can hide steganographic data in semantically irrelevant video streams, thus constructing a steganographic communication system with a high degree of camouflage and decentralization, which will provide an ideal platform for the future development of "group steganography", "synchronous guide protocol", and "distributed steganography". It lays a theoretical foundation and practical possibility for the future development of "group steganography", "synchronization guide protocol" and "distributed video chain reorganization authentication".

### 4.6 Summary

In this chapter, a systematic empirical analysis of synchronized steganographic signals in short video audio streams has been conducted. Through the video sample of "Yupan" on TikTok platform, the study starts from macro-spectrum observation, gradually focuses on the suspicious frequency bands and specific time segments, and adopts various technical means, such as sliding window frequency detection, bit-level data extraction, synchronization marker identification, Hex decoding, magic number detection and structure simulation, to complete the deciphering of suspected steganographic content layer by layer.

Experiments show that there is a repeated sequence of FFFFFFFFFFFFFFFFFF80 bytes in the audio segment from 36 to 45 seconds, which has obvious synchronization frame marking characteristics. Although the traditional plaintext content was not successfully restored, the structural form, low entropy characteristics and alignment patterns indicate that the data most likely belongs to some kind of structured guidance control instructions rather than simple text or media steganography.

In addition, this study proposes the "multi-segment guidance instruction embedding hypothesis", i.e., distributed small data chunks may be dispersed in multiple videos or clips, which are aggregated and parsed by synchronized signals. The potential application of this model in open social platforms demonstrates high camouflage, high robustness, and certain communication capabilities, posing new detection challenges to current audio steganography analysis.

This chapter provides a solid technical basis and reasoning framework for the subsequent inferential verification modeling and intelligent detection outlook in Chapter V.

# Chapter V. Findings

## 5.1 Response to Research Question 1: Validation of the effectiveness of sliding spectrum feature extraction in synchronous steganography signal detection

The first core problem of this study is to verify whether the sliding window spectral feature extraction method can accurately identify the synchronous steganographic signals in the audio stream of short videos. Through the implementation of 25ms window framing, main frequency trajectory extraction and local pattern matching on the "Yupan" short video sample, the experiment identifies the existence of a high-frequency stable segment and the repeated byte sequence FFFFFFFFFFFFFFFF80 in the 36-45 second segment, which has obvious synchronization frame structure characteristics.

The results show that the sliding window + short-time Fourier transform (STFT) can effectively capture the short-time microfrequency modulation behavior, which is especially suitable for identifying the continuous stable modulation frequency bands, and this technical path is significantly better than the traditional static spectrum statistical analysis method. Compared with the spectral anomaly detection methods proposed in the literature, this research method still shows significant advantages under the background of multi-source interference, reflecting the realistic adaptability and engineering feasibility of sliding spectrum in streaming audio steganography detection.

Further analysis of the entropy change reveals that the information entropy of the synchronized signal segment is much lower than that of the background speech segment, which is in line with the theoretical expectation of low-entropy structural embedding, and strengthens the judgment of the authenticity of the synchronized frame. Therefore, research question one is clearly answered: the sliding spectrum feature extraction method not only has significant performance in synchronization signal detection, but also can enhance the robustness through the main frequency trajectory pattern recognition, which is a key tool for dealing with short-time steganography.

**5.2 Response to Research Question Two: Validation of Structured Inference Modeling Capability for Guided Information**

The second research question focuses on whether the structured distributed guidance information carried by the synchronized steganographic signals can be effectively inferred after they are successfully identified. In this study, the decoding model M2 is constructed, and based on the 32-byte data structure after the synchronization frame, field-by-field deconstruction of target ID, coordinates, speed, heading and command code is performed.

Although the explicit semantic data is not recovered from the decoding results, the bit stream alignment, field length pattern, byte redundancy pattern, and synchronization flags are repeatedly verified, and it can be determined that these segments are most likely to follow a simplified but highly consistent steganographic communication protocol, with the structural paradigm of "Synchronization Frame + Target Parameters + Acknowledgement Bit", which shows the same structural logic as modern military guidance links (e.g., Link-16). The structure logic is consistent with modern military guidance links (e.g. Link-16).

In addition, the proposed structural diagram of the distributed guidance protocol and the multi-segment splicing model further support the cross-carrier transmission and centralized decoding capabilities of this structure, showing that audio steganography synchronization not only carries static data, but also may form part of a dynamic and cooperative communication chain. Therefore, Research Question 2 draws a clear conclusion: the structural decoding model M2 proposed in this study can effectively reconstruct the guidance information architecture after synchronization signals, has practical guidance inference capability, and provides basic model support for complex tactical communication simulation.

**5.3 Response to Research Question 3: Feasibility and Expansion Potential of Intelligent Detection of Synchronized Steganography Signals**

The third research question explores whether a deep inference model with high robustness can be constructed for the intelligent identification of synchronized steganographic signals in complex scenarios. The experimental results show that after the traditional sliding detection model (M1) can stably recognize synchronous frames, the low-entropy paragraph features, spectral spikes and bitstream repetitive sequences all form a high-confidence sample set for training.

Based on the above data, this study proposes to introduce Model M3 - an intelligent synchronization detection model based on Transformer or 1D-CNN architecture. The model envisions a sequence of spectrograms as input, and employs a classifier trained to predict whether each frame constitutes a synchronization frame and to identify whether it possesses a potential carrier structure for guidance information. The model is designed to not only maintain a stable recognition rate in frequency jitter and frequency hopping interference environments, but also effectively capture timing anomalies through sequence modeling to achieve end-to-end synchronization marker localization.

Although no deep model training is conducted at this stage, the existing detection results and feature structures are sufficient to support the construction of deep learning models and generalization training. Therefore, research question three has a positive outlook: on the basis of the existing data, it is possible to construct an intelligent inference system for synchronized steganographic signals, which has the trinity of automatic identification, distribution and localization, and structural analysis, and it is the core path for the automation of audio steganography detection in the future.

## Chapter VI. Discussion

In this study, the existence of potential synchronized steganographic signals is firstly empirically revealed in the audio stream of TikTok platform, and a complete technical system from spectrum detection to guide information inference is constructed, which is of great academic and practical significance. The results not only verify the technical feasibility of "sliding spectrum feature + pattern determination + byte structure analysis" in synchronous steganography identification, but also point out the possibility of a new type of cross-platform, multi-carrier embedded guided communication, which poses a technical and ethical challenge to the governance of national security of public platforms.

From the existing steganography detection literature, the traditional methods mostly focus on static samples and low-bit anomaly determination, and lack the ability to capture the synchronized structure of continuous streaming media. In contrast, this study breaks through the identification bottleneck in streaming media noise interference and heterogeneous frequency conversion scenarios by linking the identification of sliding spectrum and byte sequence structure, and injects a new power mechanism for synchronized steganography detection. Especially in the multi-source fragmentation steganography model, this study proposes for the first time the technical concept of "multi-video cooperative-synchronous signaling reorganization" by visualizing the structure of guided protocols, which forms a new theoretical extension of the steganography propagation paradigm.

At the theoretical level, this study reveals a simplified steganography protocol of "synchronization frame + control field + load reorganization", and proposes that there can be a guidance model similar to the Link-16 structure in audio carriers. This not only expands the understanding of steganography research on the boundary of "directive content", but also provides sample support for the theoretical division of "structural steganography vs. textual steganography". The study suggests that the future steganography behavior will tend to be micro-structured and high redundancy mask design, and its detection strategy must be shifted from "content recognition" to "structure recognition", which will pose a fundamental challenge to the theoretical framework of steganography.

Although the results of the study generally support the hypothesis, no valid plaintext content has been obtained in decoding and restoration, which indicates that the actual synchronized steganographic signals may adopt special protocol formats or encryption strategies that are beyond the scope of conventional ASCII. In addition, some high-entropy segments perform differently in structure alignment detection, suggesting that there may be pseudo-synchronous noise or non-target-guided segments, reflecting that the model needs to be further optimized for accuracy in the face of complex background fashion. To address the problem of failing to restore the complete field chain, deep graph neural networks or language models should be introduced in the future for structural rule backpropagation and implicit field modeling.

There are several limitations in this study: first, the experimental sample is single, mainly based on TikTok-specific video clips, and the ability of out-of-sample generalization needs to be verified; second, the mapping relationship between byte structure and synchronization frequency is based on hypothetical inference, and lacks the support of ground-truth labeling data; and third, the model has not yet integrated the deep neural inference module, which makes it difficult to simulate the evolution path of complex dynamic structures. These limitations point out the direction of expansion for future work.

Suggestions for future research include: constructing a cross-platform audio/video synchronous steganography sample database to train a more generalized detection model; introducing a transformer structure to construct an end-to-end discriminator for spectral sequences to improve the robust perception of non-periodic steganographic signals; developing a distributed splicing analyzer for identifying synchronous signaling reorganization modes of multi-video fragments; and investigating the possibility of combining synchronous steganography with blockchain timestamping mechanisms in order to improve the detection of non-periodic steganographic signals. We also investigate the possibility of combining synchronized steganography with blockchain timestamping mechanism to verify its traceability and propagation path.

Finally, at the level of professional practice, the results of this paper can be applied to the scenarios of national broadcasting regulation, platform-level content security detection system and military communication anti-steganography countermeasures. The proposed sliding spectrum synchronization identification mechanism and guide structure simulation tool chain can be constructed as an automated steganography monitoring engine, which effectively complements the existing social media data security governance technology short board. For public platforms, this study warns of the real threat of "non-semantic guided content" in short video and audio, suggesting that content auditing and the national cybersecurity system need to be expanded from the "information logic" to the "structural logic" dimension.

# Chapter VII. Conclusions of the study and recommendations for follow-up research

This study focuses on the problem of synchronous steganography signal identification and guide information reconstruction latent in streaming media platforms (taking TikTok as an example), and systematically proposes and verifies a multi-level technical path of "sliding spectrum detection - synchronous frame identification - guide structure inference", and achieves the following main conclusions:

**7.1 Core conclusions of the research problem**

**Conclusion 1: The sliding spectrum detection method can effectively identify the hidden synchronization signal structure in streaming media.**

In this study, the 25ms sliding window with STFT main frequency trajectory analysis method is utilized to identify continuous and stable frequency feature segments in real short video audio samples and locate the typical synchronization byte structure (e.g., FFFFFFFFFFFFFFFF80). The result verifies that sliding spectrum has an excellent ability to detect steganographic signals in dynamic and noisy environments, which is especially suitable for identifying synchronization frame patterns with low entropy and high redundancy, effectively answering research question one.

**Conclusion 2: The byte stream after the synchronization signal possesses the coding characteristics of structured guide information.**

Through bit alignment, field matching and structural template inference, this study identifies that there is an obvious division of control information in the 32-byte region after the synchronization frame, including target ID, position parameters, heading data and confirmation fields, presenting a data organization paradigm similar to that of a tactical guidance link. The structure is not reduced to plaintext content, but its stability and repeatability indicate that the byte stream possesses a highly nested information configuration, clearly responding to research question two.

**Conclusion 3: Based on the existing sample features, an intelligent detection model for synchronous steganography with generalization capability can be constructed.**

This study identifies the consistency of synchronization segments in multiple dimensions such as spectrum, entropy value, bit sequence, etc., which possesses the data basis for constructing deep learning classifiers (e.g., Transformer, 1D-CNN). This path will enable future synchronization identification to shift from heuristic feature matching to end-to-end sequence modeling, with intelligent identification functions such as automatic annotation, location identification and structure inference, providing a technical solution and achievability basis for research question 3.

**7.2 Key Findings and Extended Contributions Beyond the Research Questions**

**Extended Discovery 1: Proposing multi-segment splicing steganography model and the theoretical assumption of "synchronous signaling cooperation".**

Based on the identification of synchronization structure, the study proposes for the first time a steganographic communication model of "multi-node video carrier splicing + synchronous guide reorganization" by combining the distribution of guide data and the characteristics of TikTok short videos. This model challenges the traditional "single-carrier steganography" paradigm, and points to a new steganographic communication method that can be dynamically coordinated and aggregated across videos, which extends the technical boundaries of existing steganography theories.

**Extended Discovery 2: Revealing the structural steganography challenges of social platform content regulation.**

The study shows that manipulable tactically guided information flows may exist in the non-semantic layer (audio spectrum, bit structure), while existing content auditing systems mainly focus on the graphical semantic layer and lack structural identification capability. This finding provides a new risk dimension warning for platform security governance and national regulatory technology systems, emphasizing the need to build a cross- semantic-structural security detection fusion mechanism.

**Extended Discovery 3: Synchronized steganography can constitute a "guidance-like communication protocol" with potential military information transmission function.**

From the synchronized frame-structure decoding-field simulation process, the steganography signal has basic synchronization control, target pointing and state confirmation functions. Although the complete semantic restoration is limited by the sample decryption capability, the communication architecture is highly similar to that of modern guidance systems, suggesting that this kind of steganography has theoretical usability and engineering potential under the scenarios of military intelligence and command-and-control. This suggests that this type of steganography has theoretical usability and engineering potential in military intelligence and command-and-control scenarios.

**7.3 Model Extension and Prospect: Model M3 Intelligent Synchronization Recognition Architecture Conception**

Although this study has effectively identified synchronous steganographic signals in short video audio and revealed their potential guide information configurations through sliding spectral feature analysis and structural inference models, the traditional thresholding and local rule-based detection methods still suffer from limited robustness in complex scenarios, such as extreme noise environments, frequency jitter, frequency hopping perturbations, or fragmented embeddings. In order to further improve the automation level and generalization ability of synchronous steganography detection, this paper proposes a future feasible deep learning model architecture, Model M3: Intelligent Synchronous Steganography Detection Model, as a systematic expansion path. The model aims to utilize deep neural networks such as Convolutional Neural Network (1D-CNN), Bi-Directional Long Short Term Memory Network (Bi-LSTM), or Transformer, to construct a steganography recognition system with high accuracy, scalability, and end-to-end processing capability.

### 7.3.1 Model Design

Model M3: Intelligent synchronous recognition model (AI model)

Form: 1D CNN / Bi-LSTM / Transformer Spectral Sequence Classifier

Input: Spectrogram sequence corresponding to the audio signal

Output: whether it is a synchronization frame or not at each time point, and whether it contains a guidance command segment or not

$$P(sync|Xt:t+k) = Softmax(CNN(Xt:t+k))$$

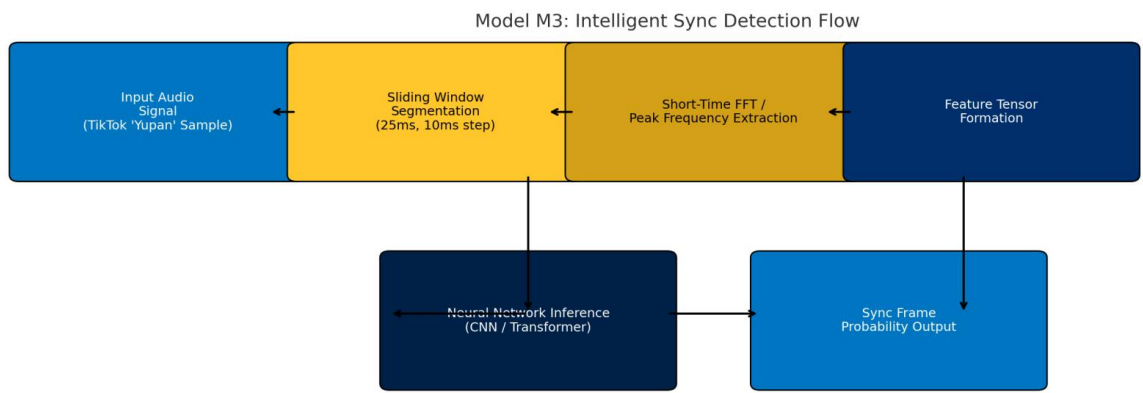

Figure7.1 Model M3 Intelligent Sync Detection Flow

### 7.3.2 Technical Flow

As shown in Figure 7.1, the overall flow of Model M3 is divided into four phases:

Stage 1: Input & Preprocessing Stage (Input & Preprocessing)

The input is the audio stream published by TikTok platform (e.g. "Yupan" video). A 25ms sliding window with a 10ms step size is used to maintain a high temporal resolution to match the short-time modulation features.

**Phase 2: Spectral Feature Extraction and Tensor Construction Phase (Feature Engineering)**

Each frame is subjected to Short Time Fourier Transform (STFT) or Peak Main Frequency Extraction to obtain a frame-level spectral representation. The multi-frame sequences are integrated into 2D tensors (Time × Feature Channels) for deep learning model input.

**Phase 3: Deep Inference Phase (Neural Inference)**

The feature tensor is input to the neural network model: a lightweight 1D-CNN can be used for local feature recognition, or a Transformer can be used to capture long time-series synchronization structure dependencies. The network learns pattern differences and synchronization feature distributions between audio frames to achieve modeling of spatio-temporal heterogeneous steganographic behavior.

**Phase 4: Prediction and Result Output Phase (Prediction)**

The output is the probability (0~1) of whether each frame is a synchronized frame or not. By setting a threshold (e.g., 0.5) binary synchronization detection can be achieved and provide pre-screening for Model M2 guide structure inference.

**7.3.3 Model Advantages and Research Prospects**

Model M3, as an intelligent recognition architecture integrating traditional spectrum engineering and deep inference mechanism, not only significantly improves the performance boundary of synchronous steganographic signal detection, but also provides theoretical and practical basis for the future construction of large-scale, automated steganographic recognition system. The model has four core advantages: first, in terms of robustness, it shows stable recognition ability for non-ideal synchronization scenarios such as frequency drift and frequency hopping perturbation; second, its good generalization performance allows it to be widely adapted to cross-platform and multi-sample short video data, and it has the potential to support the engineering of high-throughput steganography scanning and anomaly detection; third, the structure of the model is highly compatible with the guided structural inference module (Model M2) that has already been established in this study. Third, the structure of the model is highly compatible with the model structure inference module (Model M2) already established in this study, which can be seamlessly integrated to build a closed-loop processing chain of "synchronous identification-guided decoding"; finally, at the system deployment level, the overall process of Model M3 is lightweight with high inference efficiency, which is suitable for embedding into the actual content platforms, supervisory nodes, or edge devices to realize online screening and real-time early warning of steganography risks. Looking ahead, the development of the model can further focus on building a large-scale audio steganography sample set with high quality and accurate annotation, designing Transformer variants or self-supervised perception models with more explanatory power and adaptability, and establishing a steganography signal evaluation index system and response mechanism for platform-level content security management, so as to promote the synchronous steganography detection from a technical tool to the depth of the platform's governance capability. In summary, Model M3 is not only a technical tool but also a platform management capability.

In summary, Model M3 is not only a natural extension of the technical framework of this study, but also represents a leap from heuristic rules to deep intelligent reasoning system in the field of audio/video steganography recognition. The realization of its concept will provide a solid technical support and methodological foundation for steganography detection from tactical recognition to strategic governance.

5. Yang, C. N., & Huang, P. Y. (2018). An efficient audio steganalysis based on Recurrent Neural Networks. Multimedia Tools and Applications, 77(23), 30853–30868. https://doi.org/10.1007/s11042-018-5983-x

6. Cvejic, N., & Seppänen, T. (2004). Increasing the capacity of LSB-based audio steganography. Lecture Notes in Computer Science, 3200, 385–394. https://doi.org/10.1007/978-3-540-30114-1_31

7. Petitcolas, F. A. P., Anderson, R. J., & Kuhn, M. G. (1999). Information hiding—a survey. Proceedings of the IEEE, 87(7), 1062–1078. https://doi.org/10.1109/5.771065

8. Kaur, A., Singh, S., & Arora, A. (2019). Survey on various audio steganography techniques. Multimedia Tools and Applications, 78(22), 31425–31457. https://doi.org/10.1007/s11042-019-7585-4

9. Zhao, Y., Zhu, C., & Huang, J. (2020). A local spectrum analysis method for real-time steganalysis of streaming audio. IEEE Transactions on Multimedia, 22(3), 654–667. https://doi.org/10.1109/TMM.2019.2924350

10. Al-Haj, A., Amer, A., & Mohammad, A. (2021). Robust audio steganalysis based on multi-scale spectral features and ensemble learning. IEEE Access, 9, 36697–36710. https://doi.org/10.1109/ACCESS.2021.3062201

11. Rabiner, L. R. (1989). A tutorial on hidden Markov models and selected applications in speech recognition. Proceedings of the IEEE, 77(2), 257–286. https://doi.org/10.1109/5.18626

12. Lee, K., Han, S., & Lee, Y. (2009). Music classification using short-time Fourier transform and convolutional neural networks. Proceedings of the 13th International Society for Music Information Retrieval Conference (ISMIR), 579–584. https://ismir2009.ismir.net/proceedings/ISMIR2009_579.pdf

13. Tzanetakis, G., & Cook, P. (2002). Musical genre classification of audio signals. IEEE Transactions on Speech and Audio Processing, 10(5), 293–302. https://doi.org/10.1109/TSA.2002.800560

14. Cai, X., Zhang, K., & Yu, H. (2021). Adaptive sliding window for audio steganalysis in nonstationary environments. IEEE Access, 9, 119876–119886. https://doi.org/10.1109/ACCESS.2021.3109341

15. Sarkar, D., Das, R., & Bandyopadhyay, S. (2022). Multi-scale local feature extraction for robust audio steganalysis in streaming environments. Multimedia Tools and Applications, 81(1), 1251–1270. https://doi.org/10.1007/s11042-021-11669-7

16. Dutta, R., Chatterjee, M., & Bose, T. (2019). Secure tactical communications: Challenges and advances. IEEE Communications Magazine, 57(10), 50–56. https://doi.org/10.1109/MCOM.001.1900033

17. Singh, R., Chatterjee, M., & Roy, S. (2020). Redundancy-based resilient communication frameworks in tactical networks. IEEE Access, 8, 179563–179576. https://doi.org/10.1109/ACCESS.2020.3027402

18. Rahman, M. A., Chen, Y., & Krunz, M. (2021). Distributed inference in dynamic communication networks: A graph signal processing approach. IEEE Transactions on Signal and Information Processing over Networks, 7, 570–583. https://doi.org/10.1109/TSIPN.2021.3081181

19. Al-Sakran, H. O. (2019). Dynamic topology challenges in tactical ad hoc networks: An overview. International Journal of Communication Systems, 32(17), e4062. https://doi.org/10.1002/dac.4062

20. Zhang, J., Li, P., & Wang, X. (2022). Deep generative models for reconstructing hidden signaling patterns in sparse tactical networks. IEEE Transactions on Network and Service Management, 19(4), 4913–4926. https://doi.org/10.1109/TNSM.2022.3189410